\documentclass[pra,aps,superscriptaddress,showpacs,twocolumn]{revtex4}

\usepackage{amsmath,amssymb,amsthm}

\usepackage{amsgen}

\newcommand{\ra}{\rangle}

\newcommand{\id}{\mathrm{id}}
\newtheorem*{Theorem}{Theorem}

\begin{document}

\date{\today}

\title{Entanglement Equivalence of $N$-qubit Symmetric States}

\author{P. Mathonet}
\affiliation{Institut de Math\'ematique, Universit\'e de Li\`ege, 4000 Li\`ege, Belgium}

\author{S. Krins}
\affiliation{Institut de Physique Nucl\'eaire, Atomique et de
Spectroscopie, Universit\'e de Li\`ege, 4000 Li\`ege, Belgium}

\author{M. Godefroid}
\affiliation{Chimie quantique et Photophysique, CP160/09, Universit\'e Libre de Bruxelles, 1050 Bruxelles,
Belgium}

\author{L. Lamata}
\affiliation{Max-Planck-Institut f\"ur Quantenoptik,
Hans-Kopfermann-Strasse 1, 85748 Garching, Germany}

\author{E. Solano}
\affiliation{Departamento de Qu\'{\i}mica F\'{\i}sica, Universidad del Pa\'{\i}s Vasco - Euskal Herriko Unibertsitatea, Apdo.\ 644, 48080 Bilbao, Spain}
\affiliation{IKERBASQUE, Basque Foundation for Science, Alameda Urquijo 36, 48011 Bilbao, Spain}

\author{T. Bastin}
\affiliation{Institut de Physique Nucl\'eaire, Atomique et de
Spectroscopie, Universit\'e de Li\`ege, 4000 Li\`ege, Belgium}

\begin{abstract}
We study the interconversion of multipartite symmetric $N$-qubit states under stochastic local operations and classical communication (SLOCC). We demonstrate that if two symmetric states can be connected with a nonsymmetric invertible local operation (ILO), then they belong necessarily to the separable, W, or GHZ entanglement class, establishing a practical method of discriminating subsets of entanglement classes. Furthermore, we prove that there always exists a symmetric ILO connecting any pair of symmetric $N$-qubit states equivalent under SLOCC, simplifying the requirements for experimental implementations of local interconversion of those states.
\end{abstract}


\maketitle

Entanglement is a fundamental property of compound quantum systems allowing nonlocal correlations~\cite{Nielsen}. Entangled states can be further analyzed via different classification schemes. In quantum information theory, equivalence relations between entangled states are established with respect to their equivalence to implement given tasks, not necessarily with the same probability of success. This is termed as \emph{stochastic entanglement equivalence} and gives rise to the state classification under stochastic local operations and classical communication (SLOCC)~\cite{Ben00, Dur00}.
In the last years, there has been a strong boost in the study of entanglement properties of multipartite systems~\cite{Dur00,Ver02,Miy03,Ost05,Lam07,Che06,Li07,Cao07} and, in particular, of symmetric multipartite states~\cite{Tot09}. Recently, a particular interest has been raised about the possibility of classifying multipartite entanglement within the SLOCC equivalence framework from an operational point of view, be for 3 qubits~\cite{Bas09a} or for an arbitrary number of qubits in the symmetric subspace~\cite{Bas09b}. Experimental setups have also been designed to observe entire families of photon entangled states~\cite{Wie08}.

In this context, an important issue is to obtain a comprehensive identification of SLOCC equivalent classes for arbitrary number of qubits. To date this question remains unsolved as its complexity increases rapidly with the number of parties. For $N$-qubit systems, SLOCC classifications are only known for $N \leq 4$~\cite{Dur00,Ver02,Lam07}, though in the symmetric subspace the problem is solved in the general $N$ case~\cite{Bas09b}. For $N = 3$, genuine entangled states split into two inequivalent classes~: Greenberger-Horne-Zeilinger (GHZ) and W~\cite{Dur00}. For $N > 3$, the number of SLOCC classes is infinite, though they can be grouped in a finite number of families~\cite{Ver02,Lam07} and where inequivalent generalized GHZ and W classes are defined~\cite{Li07}. If the qubit system is denoted by $\{|0\rangle, |1\rangle\}$, the W class gathers states SLOCC equivalent to the paradigmatic W state
\begin{equation}
    |\mathrm{W}_N\rangle = \frac{1}{\sqrt{N}}(|10\ldots0\rangle + |010\ldots0\rangle + \cdots + |0\ldots 01\rangle),
\end{equation}
while the GHZ class gathers states SLOCC equivalent to the GHZ state
\begin{equation}
    |\mathrm{GHZ}_N\rangle = \frac{1}{\sqrt{2}}(|0 \ldots 0\rangle + |1 \ldots 1\rangle).
\end{equation}
All separable states are SLOCC equivalent to the $|0\ldots 0\ra$ state. Formally two $N$-partite states belong to the same SLOCC entanglement class if and only if they are connected via an invertible local operation (ILO)~\cite{Dur00}:
\begin{equation}
    |\psi\rangle, |\phi\rangle \in \textrm{ same class } \:\: \Leftrightarrow \:\: |\psi\rangle = A_1 \otimes \cdots \otimes A_N |\phi\rangle,
\end{equation}
where $A_i$ ($i = 1, \ldots, N$) are invertible operators ($\det A_i
\neq 0$). For arbitrary $N$, the search of ILO's connecting two states to check their SLOCC equivalence can be a formidable task, though it can be simplified by restricting the analysis to smaller subsets of states. In this paper, we show that for the important case of symmetric $N$-qubit systems it is sufficient to look for symmetric ILO's to check their SLOCC equivalence: if $|\psi_S\rangle$ and $|\phi_S\rangle$ are two symmetric $N$-qubit states, we have that
\begin{equation}
\label{symSLOCC}
    |\psi_S\rangle, |\phi_S\rangle \in \textrm{\small\ same class } \:\: \Leftrightarrow \:\: |\psi_S\rangle = A \otimes \cdots \otimes A |\phi_S\rangle,
\end{equation}
with the \emph{same} invertible operator $A$ acting on each qubit. This statement is far from being obvious, as it is easy to find many nonsymmetric ILO's connecting SLOCC equivalent symmetric states. For example, we have in the computational basis $\{|0\rangle, |1\rangle\}$,
\begin{equation}
\label{ex1}
\left(\begin{array}{cc}a_1&a_2\\a_3&a_4\end{array}\right) \otimes \left(\begin{array}{cc}a_1&b_2\\a_3&b_4\end{array}\right) |00\rangle = |\phi \phi\rangle,
\end{equation}
with
\begin{equation}
|\phi\rangle = a_1 |0\rangle + a_3 |1\rangle,
\end{equation}
and where $a_i$ and $b_i$ ($i = 1, \ldots, 4$) are complex numbers. The SLOCC transformation of Eq.~(\ref{ex1}) is clearly not symmetric, though the two separable states $|00\rangle$ and $|\phi \phi \rangle$ are. This trivial example shows that two SLOCC equivalent symmetric states can be connected by nonsymmetric ILO's. This observation is not restricted to the case of separable states. For instance, one checks straightforwardly that the GHZ class state
\begin{equation}
\label{ex2}
|\psi\rangle = (A \otimes B \otimes C) \frac{|000\rangle + |111\rangle}{\sqrt{2}}
\end{equation}
is symmetric as long as the 3 matrices $A B^T$, $B C^T$ and $C A^T$ are themselves all symmetric. Many distinct (and nonproportional) invertible matrices $A$, $B$ and $C$ can satisfy those 3 conditions, like, e.g.,
\begin{equation}
    A = \left(\begin{array}{cc}a &c\\b&0\end{array}\right), \quad B = \left(\begin{array}{cc}a &d\\b&0\end{array}\right), \quad C = \left(\begin{array}{cc}a &e\\b&0\end{array}\right),
\end{equation}
with $a, b, c, d$ and $e$ nonzero complex numbers.

These two examples show clearly that many nonsymmetric ILO's can connect symmetric states. It is therefore an important issue to know whether the analysis of the SLOCC equivalence of symmetric states can be restricted to the search of only symmetric ILO's connecting them. We prove hereafter that this is the case and that Eq.~(\ref{symSLOCC}) strictly holds. This represents an important simplification in the analysis of SLOCC equivalence of symmetric states. In the light of both examples given above, this fundamental result can be restated as follows: \emph{If a nonsymmetric ILO connects two symmetric states $|\psi\rangle$ and $|\phi\rangle$, then it always exists a symmetric ILO connecting them}. Furthermore, we will also show that the occurrence of nonsymmetric ILO's between symmetric states only exists for separable, W and GHZ class states. For $N \geq 4$, where the number of classes is infinite, these reduced subsets of states represent an infinitesimal fraction of SLOCC classes. More formally our result reads
\begin{Theorem}\label{mainT}
If $|\psi\ra$ and $|\phi\ra$ are symmetric $N$-qubit states, and if there exist invertible $A_1,\ldots, A_N$ operators such that
\begin{equation}
\label{SLOCCthm}
|\phi\ra = A_1\otimes\cdots\otimes A_N|\psi\ra,
\end{equation}
then there necessarily exists an invertible operator $A$ such that
\begin{equation}
\label{SSLOCCthm}
|\phi\ra = A^{\otimes N}|\psi\ra.
\end{equation}
\end{Theorem}

When all invertible operators $A_i$ are proportional to each other without being all identical, the theorem is trivial. In this case $A_1\otimes\cdots\otimes A_N = (c A_1)^{\otimes N}$ with $c$ an $N$-th root of the product of all proportionality constants of the invertible operators with respect to $A_1$. We focus hereafter only on the nontrivial case where at least two operators are not proportional to each other, say $A_i$ and $A_j$ ($1 \leqslant i,j \leqslant N$). We can first observe that, as $|\psi\ra$ and $|\phi\ra$ are symmetric with respect to permutations of the qubits, we have
\begin{equation}
\label{sigma}
\sigma |\psi\ra = |\psi\ra, \quad \sigma |\phi\ra = |\phi\ra
\end{equation}
for any permutation $\sigma$ of $\{1, \ldots, N \}$. Inserting these 2 equalities into Eq.~(\ref{SLOCCthm}) yields
\begin{equation}
|\phi\ra = \sigma^{-1} (A_1\otimes\cdots\otimes A_N) \sigma |\psi\ra,
\end{equation}
that is
\begin{equation}
|\phi\ra = A_{\sigma(1)}\otimes\cdots\otimes A_{\sigma(N)} |\psi\ra.
\end{equation}
This shows that the ordering of the operators $A_1, \ldots, A_N$ in the tensorial product in Eq.~(\ref{SLOCCthm}) is completely arbitrary and can be changed at will. We use this property to set the operators $A_i$ and $A_j$ in the first two positions. We have if $\mu$ is any numbering of $\{1,\ldots,N\}\setminus\{i,j\}$
\begin{equation}
|\phi\ra = A_i\otimes A_j\otimes A_{\mu_1}\otimes\cdots\otimes A_{\mu_{N-2}}|\psi\ra
\end{equation}
and
\begin{equation}
|\phi\ra = A_j\otimes A_i\otimes A_{\mu_1}\otimes\cdots\otimes A_{\mu_{N-2}}|\psi\ra.
\end{equation}
Therefore, we get
\begin{equation}
\label{inv1}
A_i^{-1}A_j\otimes A_j^{-1}A_i\otimes {\id}^{\otimes N-2}|\psi\ra=|\psi\ra,
\end{equation}
where $\id$ denotes the identity operator. Eq.~(\ref{inv1}) shows that the state $|\psi\ra$ of Eq.~(\ref{SLOCCthm}) is invariant under the action of the operation $A_i^{-1}A_j\otimes A_j^{-1}A_i\otimes {\id}^{\otimes N-2}$.

In any representation the invertible matrix $A_i^{-1}A_j$ can be reduced via a similarity transformation either to a diagonal matrix (case 1)
\begin{equation}
\label{D}
D = \left(\begin{array}{cc}\lambda_1&0\\0&\lambda_2\end{array}\right),
\end{equation}
with nonzero $\lambda_1$ and $\lambda_2$ coefficients,
either to a true Jordan matrix (case 2)
\begin{equation}
J = \left(\begin{array}{cc}\lambda&1\\0&\lambda\end{array}\right),
\end{equation}
with a nonzero $\lambda$ coefficient,
the later case occurring only when the eigenvalues of $A_i^{-1}A_j$ are identical and $A_i^{-1}A_j$ is not diagonalizable.
Hereafter, we consider both cases and work in the computational basis $\{|0\rangle, |1\rangle\}$.

\textit{Case 1.} In this case we know that it exists an invertible matrix $S$ and a diagonal matrix $D = \mathrm{diag}(\lambda_1, \lambda_2)$ such that
\begin{equation}
A_i^{-1}A_j=SDS^{-1}.
\end{equation}
The two nonzero diagonal elements $\lambda_1$ and $\lambda_2$ are distinct, otherwise $D$ would be a multiple of the identity matrix and $A_j$ would be proportional to $A_i$, which would contradict our starting assumption on those two matrices.
We can then rewrite relation (\ref{inv1}) as
\begin{equation}
SDS^{-1}\otimes SD^{-1}S^{-1}\otimes (SS^{-1})^{\otimes N-2}|\psi\ra=|\psi\ra,
\end{equation}
and setting
\begin{equation}
\label{psippsi}
|\psi'\ra=S^{{-1}^{\otimes N}}|\psi\ra,
\end{equation}
we get
\begin{equation}\label{strong}
D\otimes D^{-1}\otimes \id^{\otimes N-2}|\psi'\ra=|\psi'\ra.
\end{equation}

Next, we develop the symmetric state $|\psi'\ra$ over the unnormalized symmetric Dicke state basis:
\begin{equation}\label{basis}
|\psi'\ra=\sum_{k=0}^Na_k|\psi_N^{(k)}\ra,
\end{equation}
where
\begin{equation}
|\psi_N^{(k)}\ra = \sqrt{C_N^k} |D_N^{(k)}\ra, \quad k = 0, \ldots, N .
\end{equation}
Here, $C_N^k$ are the binomial coefficient of $N$ and $k$, and $|D_N^{(k)}\ra$ the normalized Dicke state
\begin{equation}
|D_N^{(k)}\ra = \frac{1}{\sqrt{C_N^k}} \sum_{P(0,1)} |\underbrace{0\cdots0}_{N-k}\underbrace{1\cdots1}_k\ra, \quad k = 0, \ldots N,
\end{equation}
where the sum runs over all  $C_N^k$ permutations of 0 and 1 appearing $N-k$ and $k$ times, respectively. The Dicke state $|D_N^{(1)}\ra$ identifies to the W state $|\mathrm{W}_N\ra$. For $N \geqslant 3$ and using the convention $|\psi_N^{(k)}\rangle = 0$ for $k < 0$ or $k > N$, the unnormalized symmetric Dicke states are linked by the relation
\begin{align}
\label{relpsiNk}
|\psi_N^{(k)}\ra & = |00\ra\otimes|\psi_{N-2}^{(k)}\ra+(|01\ra+|10\ra)\otimes|\psi_{N-2}^{(k-1)}\ra
\nonumber \\ & ~\quad + |11\ra\otimes|\psi_{N-2}^{(k-2)}\ra.
\end{align}
This relation still holds for $N = 2$ if we formally set $|\psi_0^{(0)}\ra = 1$.

Using Eqs.~(\ref{basis}), (\ref{relpsiNk}), and the obvious identities $D\otimes D^{-1}|00\ra=|00\ra$, $D\otimes D^{-1}|11\ra=|11\ra$ and $D\otimes D^{-1}(|01\ra+|10\ra)=\frac{\lambda_1}{\lambda_2}|01\ra+\frac{\lambda_2}{\lambda_1}|10\ra$, Eq.~(\ref{strong}) simplifies to
\begin{equation}
|\psi'\ra+\sum_{k=0}^N a_k ((\frac{\lambda_1}{\lambda_2}-1)|01\ra+(\frac{\lambda_2}{\lambda_1}-1)|10\ra)\otimes |\psi_{N-2}^{(k-1)}\rangle = |\psi'\ra.
\end{equation}
Hence we must have
\begin{equation}
\label{newrel}
\left[(\frac{\lambda_1}{\lambda_2}-1)|01\ra+(\frac{\lambda_2}{\lambda_1}-1)|10\ra\right]\otimes \sum_{k=0}^Na_k|\psi_{N-2}^{(k-1)}\ra=0.
\end{equation}
As $|\psi_{N-2}^{(k-1)}\ra=0$ by convention for $k=0$ and $k=N$, and as all other states $|\psi_{N-2}^{(k-1)}\ra$ are linearly independent, Eq.~(\ref{newrel}) implies that $a_k$ vanishes for $k = 1, \ldots, N-1$ and we conclude that $|\psi'\rangle$ is necessarily of the form
\begin{align}
|\psi'\ra & = a_0|\psi_{N}^{(0)}\ra+a_N|\psi_{N}^{(N)}\ra \nonumber \\ & = a_0 |0\ldots0\rangle + a_N |1\ldots1\rangle.
\end{align}
If $a_N$ or $a_0$ vanishes, $|\psi'\ra= S'^{\otimes N}|0\ldots0\ra$ with
\begin{equation}
    \label{Spa0}
    S' = \left(\begin{array}{cc}a_0^{1/N}&0\\0&1\end{array}\right)
\end{equation}
or
\begin{equation}
    S'=\left(\begin{array}{cc}0&1\\a_N^{1/N}&0\end{array}\right),
\end{equation}
respectively.
If $a_0$ and $a_N$ are both nonzero, $|\psi'\ra = S'^{\otimes N}|\mathrm{GHZ}_N\ra$ with
\begin{equation}
S' = 2^{1/2N} \left(\begin{array}{cc}a_0^{1/N}&0\\0&a_N^{1/N}\end{array}\right).
\end{equation}
Using Eq.~(\ref{psippsi}), we conclude that when present case 1 holds, the state $|\psi\ra$ is either equal to $(SS')^{\otimes N}|0 \ldots 0\rangle$, or to $(SS')^{\otimes N}|\mathrm{GHZ}_N\rangle$, where $SS'$ is invertible.

\textit{Case 2.} In this case we know that it exists an invertible matrix $S$ such that
\begin{equation}
A_i^{-1}A_j=SJS^{-1},\quad\mbox{where}\quad J=\left(\begin{array}{cc}
 \lambda&1\\0&\lambda
\end{array}\right),
\end{equation}
with $\lambda \neq 0$.
The relation (\ref{inv1}) can be rewritten here as
\begin{equation}
SJS^{-1}\otimes SJ^{-1}S^{-1}\otimes (JJ^{-1})^{\otimes N-2}|\psi\ra=|\psi\ra.
\end{equation}
Setting again $|\psi'\ra=S^{{-1}^{\otimes N}}|\psi\ra$, we obtain the invariance relation
\begin{equation}
\label{strongJ}
J\otimes J^{-1}\otimes \id^{\otimes N-2}|\psi'\ra=|\psi'\ra.
\end{equation}
By decomposing the state $|\psi'\ra$ over the unnormalized symmetric Dicke state basis as in Eq.~(\ref{basis}), and by using Eq.~(\ref{relpsiNk}) and the obvious identities $J\otimes J^{-1}|00\ra=|00\ra$, $J\otimes J^{-1}(|01\ra+|10\ra)=|01\ra+|10\ra$ and  $J\otimes J^{-1}|11\ra=-\frac{1}{\lambda^2}|00\ra+\frac{1}{\lambda}|01\ra-\frac{1}{\lambda}|10\ra+|11\ra$, Eq.~(\ref{strongJ}) simplifies to
\begin{equation}
|\psi'\ra+\left[-\frac{1}{\lambda^2}|00\ra+\frac{1}{\lambda}|01\ra-\frac{1}{\lambda}|10\ra \right]\otimes\sum_{k=0}^Na_k|\psi_{N-2}^{(k-2)}\ra = |\psi'\ra,
\end{equation}
which implies
\begin{equation}
\label{newrel2}
\left[-\frac{1}{\lambda^2}|00\ra+\frac{1}{\lambda}|01\ra-\frac{1}{\lambda}|10\ra\right]\otimes\sum_{k=0}^Na_k|\psi_{N-2}^{(k-2)}\ra = 0.
\end{equation}
As $|\psi_{N-2}^{(k-2)}\ra=0$ by convention for $k=0$ and $k=1$, and as all other states $|\psi_{N-2}^{(k-2)}\ra$ are linearly independent,
the $a_k$ coefficients must vanish for $k = 2, \ldots, N$ and $|\psi'\rangle$ is necessarily of the form
\begin{equation}
|\psi'\ra=a_0|\psi_{N}^{(0)}\ra+a_1|\psi_{N}^{(1)}\ra.
\end{equation}
If $a_1 = 0$, we have trivially $|\psi'\ra = S'^{\otimes N}|0\ldots0\ra$ with $S'$ as given by Eq.~(\ref{Spa0}). If $a_1 \neq 0$, $|\psi'\ra = S'^{\otimes N}|\mathrm{W}_N\rangle$ with
\begin{equation}
\label{lastSp}
S'={N}^{1/2N}\left(\begin{array}{cc}1&a_0/N\\0&a_1\end{array}\right).
\end{equation}
When present case 2 holds the state $|\psi\ra$ is either equal to $(SS')^{\otimes N}|0 \ldots 0\rangle$, or to $(SS')^{\otimes N}|\mathrm{W}_N\rangle$, $SS'$ being invertible.

Consequently, whatever case holds above, we can always find, for any symmetric state $|\psi\ra$ connected via a nonsymmetric ILO to another symmetric state (Eq.~(\ref{SLOCCthm}) in the nontrivial case), an invertible local operation $S_{\psi}$ expressing the state either as $S_{\psi}^{\otimes N}|0 \ldots 0\rangle$, $S_{\psi}^{\otimes N}|\mathrm{W}_N\ra$ or $S_{\psi}^{\otimes N}|\mathrm{GHZ}_N\ra$. The operation $S_{\psi}$ is given by the product $S S'$ defined above as a function of the different cases. The state $|\psi\ra$ belongs then necessarily to the separable, W or GHZ class.

All the argumentation we have used to characterize the state $|\psi\ra$ can be repeated similarly for the state $|\phi\ra$ on the basis of the inverse of Eq.~(\ref{SLOCCthm}), namely
\begin{equation}
    |\psi\rangle = A_1^{-1} \otimes \ldots \otimes A_N^{-1} |\phi\rangle.
\end{equation}
By using the same reasoning as that from Eq.~(\ref{sigma}) to Eq.~(\ref{lastSp}) where the operators $A_{\mu}^{-1}$ ($\mu = 1, \ldots, N$) are considered instead of the $A_{\mu}$, an invertible local operation $S_{\phi}$ is shown to exist that expresses the symmetric state $|\phi\ra$ either as $S_{\phi}^{\otimes N}|0 \ldots 0\rangle$, $S_{\phi}^{\otimes N}|\mathrm{W}_N\ra$ or $S_{\phi}^{\otimes N}|\mathrm{GHZ}_N\ra$. The state $|\phi\ra$ belongs either to the separable, W, or GHZ class.

Since by our starting hypothesis the states $|\psi\ra$ and $|\phi\ra$ are SLOCC equivalent and since the states $|0 \ldots 0\rangle$, $|\mathrm{W}_N\ra$ and $|\mathrm{GHZ}_N\ra$ are all SLOCC inequivalent, $|\psi\ra$ and $|\phi\ra$ must be \emph{both} either separable, either of the W class, or of the GHZ class. We then conclude that Eq.~(\ref{SSLOCCthm}) holds for
\begin{equation}
    A = S_{\phi} S_{\psi}^{-1}.
\end{equation}

Finally, it is noteworthy to mention that these results have also important practical consequences, be for discriminating subsets of entanglement classes~\cite{Gue09} or for the experimental generation of multipartite symmetric entangled states~\cite{Sch05}. In the former, the existence of a nonsymmetric ILO between two symmetric states guarantees that those states belong to the $\{ \rm Separable, W, GHZ \}$ class subset. In the latter, we have a general simplified recipe for transforming one symmetric $N$-qubit state to another in experimental implementations, given that we know now a nontrivial statement: \emph{there always exists a symmetric ILO that connects any pair of SLOCC equivalent symmetric states}. For symmetric $N$-qubit states, the SLOCC classes are entirely spanned with help of a \emph{single} local operation $A$ acting similarly on each qubit. This represents only 4 independent parameters, whatever the number of qubits involved in the operation. This must be contrasted with the $4N$ independent parameters of a general nonsymmetric ILO.

In conclusion, we have shown in this paper that SLOCC equivalent symmetric states can always be connected through symmetric ILO's. The nonexistence of such symmetric ILO's between two states is sufficient to prove the SLOCC inequivalence of the states. We have also shown that when a nonsymmetric ILO connects two symmetric states, this can only happen when the two states are either separable, of the W or of the GHZ class. This gives those classes a very peculiar status amongst the infiniteness of the SLOCC classes when the number of qubits exceeds 3. For the latter case, we have given a protocol on how to find a symmetric ILO connecting the associated states.

The authors thank A. Osterloh for helpful discussions.
L.L. thanks the Alexander von Humboldt Foundation
for funding. E.S. acknowledges UPV-EHU Grant No. GIU07/40 and the EuroSQIP European project. T.B., S.K. and M.G. thank the Belgian FRS-FNRS.

\end{document}